\documentclass[12pt]{article}

\usepackage[super,compress]{cite}
\usepackage{graphicx}
\usepackage{comment}
\renewcommand{\d}{{\rm d}}
\newcommand{\rl}{\ell}
\newcommand{\tilS}{\tilde{S}}
\newcommand{\bN}{\mbox{\boldmath$N$}}
\newcommand{\rml}{ l}
\newcommand{\bpart}{\mbox{\boldmath$\partial$}}
\newcommand{\bn}{\mbox{\boldmath$n$}}
\newcommand{\bx}{\mbox{\boldmath$x$}}
\newcommand{\tl}{\tilde{l}}
\newcommand{\bDelta}{\overline{\Delta}}
\newcommand{\bp}{\mbox{\boldmath$p$}}
\newcommand{\vf}{\mbox{\boldmath$f$}}
\begin{document}

\title{On the discrete version of the Reissner–Nordström solution
}

\author{V.M. Khatsymovsky \\
 {\em Budker Institute of Nuclear Physics} \\ {\em of Siberian Branch Russian Academy of Sciences} \\ {\em
 Novosibirsk,
 630090,
 Russia}
\\ {\em E-mail address: khatsym@gmail.com}}
\date{}
\maketitle
\begin{abstract}
This paper generalizes our previous paper on the discrete Schwarzschild type solution in the Regge calculus, the simplicial electrodynamics earlier considered in the literature is incorporated in the case of the presence of a charge. Validity of the path integral approach is assumed, of which the only consequence used here is a loose fixation of edge lengths around a finite nonzero scale (we have considered the latter earlier).

In essence, the problem of determining the optimal background metric and electromagnetic field for the perturbative expansion generated by the functional integral is considered, for which the skeleton Regge and electrodynamic equations are analyzed. For the Regge equations, as we have earlier found, the Regge action on the simplest periodic simplicial structure and in the leading order over metric variations between 4-simplices can be substituted by a finite-difference form of the Hilbert-Einstein action (the piecewise constant metric there is defined by providing the vertices with coordinates).

Thus we get the absence of the singularity inherent in the continuous solution. At the same time, the discrete solution is close to the continuum Reissner–Nordström one at large distances.
\end{abstract}


MSC classes: 83C27; 83C57

keywords: Einstein theory of gravity; minisuperspace model; piecewise flat spacetime; Regge calculus; Reissner–Nordström black hole

\section{Introduction}

In extreme gravity, say, in a black hole, the simplicial structure of spacetime can manifest itself, if any. Here we are talking about a piecewise flat ansatz for the Riemannian manifold introduced by Regge \cite{Regge}, considered not just as a mathematical approximation to the smooth Riemannian manifold, but as a real physical system. A piecewise flat manifold can be viewed as composed of flat 4-dimensional tetrahedra or 4-simplices that make up a simplicial complex. The geometry of such a system is characterized by a countable number of edge lengths. The countability of the set of field variables is promising in view of the formal nonrenormalizability of the conventional general relativity with a continuous set of variables and contributes to the consistent definition of the functional integral \cite{Ham,Ham1}. At the same time, a Riemannian manifold can, in a certain sense, be arbitrarily accurately approximated by a piecewise flat manifold \cite{Fein,CMS}. The functional integral can be used to find physical quantities in this discrete setting \cite{HamWil1,HamWil2}. In fact, to approximate a Riemannian manifold by a piecewise flat one, several types of the 4-simplices are sufficient; confining ourselves to them leads to the Causal Dynamical Triangulations approach \cite{cdt,TaVis1,TaVis2}. Reducing the number of types of the 4-simplices makes it possible to sum over histories in the path integral by numerical methods. Using a simplicial complex here is rather a regularization method. The concept of the real simplicial structure of spacetime was used for quantizing gravity as well \cite{Mik,Mik1}.

The Regge calculus was applied to the numerical analysis of cosmological models \cite{WilCol,Gen,Bre4,WilLiu} or, being modified to the form of the Causal Dynamical Triangulations approach, to the emergence of cosmological models in quantum gravity in the framework of a non-perturbative path integral over spacetimes \cite{GlaLol}. The 3D Regge calculus (a fixed simplicial decomposition of 3D space) was applied to a numerical application to the Schwarzschild and Reissner-Nordström geometries \cite{Wong}. In what follows, more efficient computational lattice methods were proposed for such systems \cite{Bre2}. Schwarzschild's black hole was considered in Loop Quantum Gravity \cite{Ash1,Ash2}, where the central singularity was cut off at the Planck scale due to the existence of a Planck-scale area quantum.

We considered the Schwarzschild black hole in Regge calculus \cite{Kha1,our}. In fact, there we turned to solving the problem of determining the optimal background metric for the perturbative expansion generated by the functional integral. The quantum or functional integral constituent of the problem is simply reduced to the existence of some elementary length scale that maximizes the functional integral measure and around which the edge lengths are loosely fixed. So, we consider solving the Regge equations with this elementary length scale. In the classical limit, the elementary length scale tends to zero, discreteness transforms into continuity, Regge equations turn into Einstein equations with their singularities. But while we are still within the framework of quantum theory, these singularities are cut off at the elementary length scale. Now consider incorporating matter fields into the system. General aspects of incorporating matter fields into the discrete gravity were reviewed in, e. g., [\citen{Ham,Ham1}]. We are interested in the electromagnetic field and, correspondingly, the implementation of the simplicial electrodynamics considered in the literature \cite{Sor,Wein}.

This is done by example of the Reissner–Nordström black hole \cite{Reis,Nord}. The method is described in Section \ref{method}. The description of the simplicial electromagnetic field used is considered in Section \ref{em-field}. The mechanism of the loose fixation of the elementary length scale is seen upon the functional integration over connection and can be analysed with the help of the expansion of the functional integral over the discrete lapse-shift functions used as constant parameters. The latter takes place, in particular, in (a discrete analogue of) a synchronous frame of reference. In the leading order over metric variations from simplex to simplex, we can pass to (a discrete version of) any other coordinate system, for which we choose the Kerr-Schild one. In this order, the details of this intermediate synchronous frame are irrelevant, only its existence is important. This point is considered in Section \ref{ansatz}. The system of the discrete Einstein (Regge) - Maxwell equations and the solution to it is considered in Section \ref{solution}. The Riemann tensor at the center is evaluated in Section \ref{riemann}.

\section{The method}\label{method}

A triangulation of spacetime, as a measurement process, in quantum theory must have limitations in accuracy.
In particular, we can talk about the most detailed, in a certain sense, triangulation. As a result of this triangulation, we have a simplicial complex composed of 4-simplices. It is natural to assume that these 4-simplices are flat (allowing their non-trivial structure, in particular, their internal curvature, would contradict the assumption of the most detailed triangulation). This assumes a piecewise flat space-time and a probability distribution of the values of the scale of elementary lengths having a maximum at a non-zero value. We find such a distribution or the functional integral measure arising upon integrating out the connection variables \cite{our1}. In Regge calculus, it is customary to average/sum over all kinds of simplicial structures, but here we consider a certain simplest periodic structure.

The use of discrete connection variables introduced by Fr\"{o}hlich \cite{Fro} is convenient when we are trying to establish a relation to the canonical Hamiltonian formalism, which would be as less singular as possible and which could be constructed if we choose a certain coordinate as time and pass to the continuous time limit. If we write the Regge action $S_{\rm g} ( \rl )$ in a certain way in terms of both edge vectors $\rl = (l_1, \dots, l_n )$ and dependent on them connection $\Omega$ (SO(3,1) rotations defined on the tetrahedra or 3-simplices) as $S_{\rm g}(\rl , \Omega )$, then consider $\Omega$ as independent variables and exclude them via classical equations of motion, then we get exactly the original Regge action $S_{\rm g} ( \rl )$; the same holds if $S_{\rm g}(\rl , \Omega )$ is one of two such forms in which $\Omega$ are only self-dual or only anti-self-dual rotations \cite{Kha}, or a combination of both.

In these independent variables $\rl , \Omega$, the canonical path integral measure is still singular, which is connected with an abrupt change in the symmetry of the discrete theory in the vicinity of the flat background. Such a measure can be defined nonsingularly in the extended configuration superspace of {\it independent} area tensors of the triangles or 2-simplices (and thus ambiguous edge vectors). Then the full discrete measure can be constructed which in a sense tends to the canonical one in the continuous time limit, regardless of which coordinate is chosen as time. We can project this measure onto the physical hypersurface of the unambiguous edge vectors by inserting an appropriate $\delta$-function factor. As a result, we have a measure $\d \mu ( \rl ) {\cal D} \Omega$ with the edge vector part of the measure $\d \mu ( \rl )$ and the connection part of the measure ${\cal D} \Omega$. Performing functional integration over $\Omega$, we get a measure (modulus) $F ( \rl )$ and an argument (phase) $\tilS_{\rm g} ( \rl )$ of the resulting functional integral in terms of edge vectors $\rl$ only,
\begin{equation}\label{int-exp(iS)-dmu-DOmega}                              
\int \exp [ i S_{\rm g}(\rl , \Omega  ) ] ( \cdot ) \d \mu ( \rl ) {\cal D} \Omega = \int \exp [ i \tilS_{\rm g} ( \rl ) ] ( \cdot ) F ( \rl ) D \rl
\end{equation}

\noindent (it can be applied to the functions of observables $\rl$). Convenient methods for computing $F ( \rl )$ and $\tilS_{\rm g} ( \rl )$ can be based on expansions of different types.

For $F ( \rl )$, it seems appropriate to expand the LHS of (\ref{int-exp(iS)-dmu-DOmega}) over discrete analogs of the tetrad vector $e_0^a$ with the temporal world index 0 and the local frame index $a$ (discrete ADM lapse-shift functions \cite{ADM1} $(N, \bN )$), since this gives a nontrivial factor in $F ( \rl )$ already in the zero order of this expansion. Integration over $\Omega$ splits into integrations over (independent components of) the holonomy $R$ of $\Omega$ (curvature) and over a gauge group that do not change $R$. The latter results in the gauge group volume and cancels out due to the normalization of the functional integral. The integral over non-compact degrees of freedom in $R$ is improper, but conditionally convergent due to the oscillating exponent. In principle, the integral over connection can be found term by term for the considered expansion \cite{Kha3}, thus generating an expansion for $F ( \rl )$.

Also we can use the stationary phase expansion when integrating over $\Omega$. We get a nontrivial $\tilS_{\rm g} ( \rl )$ already in the zero order of this expansion. Note that we are considering the stationarity of the phase not with respect to the general set of variables $\Omega$, $\rl$, but only with respect to the connection $\Omega$. As a continuum analogue, we have the Gaussian integration over the (infinitesimal) connection $\omega \in$ so(3,1) and, as a result, exactly the usual path integral in terms of only tetrad (or metric) $e$, $\int \exp ( iS_{\rm g} ( e ) ) \d \mu ( e )$, with the phase $S_{\rm g} ( e )$ being the Einstein-Hilbert action.

In the discrete setting, infinitesimal $\omega$ appear as generators of $\Omega \in SO(3,1)$ (or, more generally, of $\Omega = \Omega_0 \exp \omega$ for a non-trivial background $\Omega_0$), and we have lattice-specific terms of the order $[\omega]^n$, $n > 2$, additional to the bilinear form of $\omega$ in the action. So $\tilS_{\rm g} ( \rl )$ is the Regge action $S_{\rm g} ( \rl )$ in the zero order (since $S_{\rm g} ( \rl )$ can be obtained from $S_{\rm g}(\rl , \Omega  )$ by substituting a solution of the equations of motion for $\Omega$ by definition of the representation of the Regge action) plus lattice-specific corrections of the order $l^{-2k}$, $k \geq 1$, where $l$ is the typical edge length. These corrections can be small if $l \gg 1$ in Planck units. In our self-regulating scheme, $l$ is defined below as $b$ (\ref{b=sqrt}), determined by the measure. Therefore, in order to be able to restrict ourselves to the Regge action for the phase, the fundamental parameter $\eta$ is assumed to be large there: $\eta \gg 1$.

One can restrict oneself to the zero order terms in the expansions for both $F ( \rl )$ and $\tilS_{\rm g} ( \rl )$, if $b \geq 1$ below, as we consider in [\citen{our1}].

Upon taking into account the electromagnetic field characterized by simplicial variables $A$ and action $S_{\rm em}$ specified below, the full action is
\begin{equation}                                                            
S(\rl , \Omega , A ) = S_{\rm g}(\rl , \Omega  ) + S_{\rm em}(\rl , A  ) .
\end{equation}

\noindent Upon integrating out $\Omega$, we have the same $\tilS_{\rm g} ( \rl )$ and $F ( \rl )$, and first concentrate on the dependence on $\rl$ considering $A$ in the effective action $S(\rl , A ) = \tilS_{\rm g}(\rl  ) + S_{\rm em}(\rl , A  ) \approx  S_{\rm g}(\rl  ) + S_{\rm em}(\rl , A  )$ as parameters.

The functional integral generates some perturbative expansion. Let us expand the phase in some new variables $u = (u_1, \dots, u_n )$, which we introduce instead of $\rl $ in order to transform the measure $F( \rl ) D \rl$ into the Lebesgue measure $D u$:
\begin{equation}\label{Sdudu}                                               
S (\rl , A ) = S (\rl_0 , A ) + \frac{1}{2} \sum_{j, k, l, m} \frac{\partial^2 S (\rl_0 , A )}{\partial l_j \partial l_l} \frac{\partial l_j (u_0 )}{\partial u_k} \frac{\partial l_l (u_0 )}{\partial u_m} \Delta u_k \Delta  u_m + \dots ,
\end{equation}

\noindent $\Delta u = u - u_0$, provided that
\begin{equation}\label{dS/dl=0}                                             
\frac{\partial S(\rl_0 , A )}{\partial l_j} = 0 .
\end{equation}

\noindent Just as we impose the extremum condition (\ref{dS/dl=0}) on the zero-order term to maximize the contribution, now we can also impose a minimum condition on the determinant of the second-order form in the exponential for this, that is, the condition for the maximum value of
\begin{equation}\label{def-l0}                                              
F (\rl_0 )^2 \det \left \| \frac{\partial^2 S (\rl_0 , A )}{\partial l_i \partial l_k} \right \|^{-1} .
\end{equation}

\noindent Eqs (\ref{dS/dl=0}), (\ref{def-l0}) define a certain optimal initial point $\rl_0 = (l_{01}, \dots, l_{0n} )$ (background metric).

If we are interested in the dependence on the length scale $\rml$ in a certain domain, then in $F ( \rl ) D \rl$, this dependence is manifested mainly in the form of a factor $[f( \rml )]^T \rml^{-1} \d \rml$, where $f$ is a certain function, $T$ is the number of certain - spatial and diagonal - triangles in this domain, as we consider in [\citen{our1}], there in terms of an area scale. (Thereby, in fact, maximizing (\ref{def-l0}) gives $\rml$ as the scale of the edge lengths of the spatial and diagonal edges, but not of the temporal ones, whose vectors, the discrete analogs of the lapse-shift functions, are fixed by hand.) $T$ is expected to be large (far from singularities; in the vicinity of a singularity, $\rml$ changes abruptly, $T \sim 1$, but there, strictly speaking, also higher orders over metric variations should be taken into account; we perform an order-of-magnitude extrapolation to the singularity from the side of small metric variations). $S_{\rm g}$ and $S_{\rm em}$ depend on $\rml$ as $\rml^2$ and $\rml^0$, respectively, so if we would substitute $S$ by $S_{\rm g}$ or $S_{\rm em}$, then taking into account $\partial^2 S (\ell_0 , A ) / \partial l_i \partial l_k$ in (\ref{def-l0}) would result in maximizing $[f( \rml )]^T \rml^{-1}$ or $[f( \rml )]^T$, respectively. That is, taking into account $\partial^2 S (\ell_0 , A ) / \partial l_i \partial l_k$ in (\ref{def-l0}) is inessential (for large $T$) for defining the optimal length scale and, in particular, taking into account $S_{\rm em}$ does not significantly affect the length scale.

Thus, the matrix $\partial^2 S (\ell_0 , A ) / \partial l_i \partial l_k$ has a relatively low, in the just considered sense, order of dependence on the edge lengths scale. Any element of this matrix is non-zero only for those $l_i$ and $l_k$ that can be assigned to the same 4-simplex and can depend, apart from some common length scale, on the ratios of the edge lengths related to this 4-simplex. Geometrically, the edge length scale can not change sharply from simplex to simplex. Therefore, we expect that taking into account the determinant of this matrix in (\ref{def-l0}) can not significantly change the point $\ell = \ell_0$ of the maximum of (\ref{def-l0}) in comparison with the maximum of $F (\ell )$ only. This also means some sufficient independence from the location where the elementary length scale is determined.

Finally, we can consider the expansion of the effective action over $A$; as usual, the requirement of the absence of linear (over variations of $\rl$, $A$) terms in the path integral phase gives the equations of motion for $A$ at a solution $A = A_0$,
\begin{equation}\label{dS/dA=0}                                             
\frac{\partial S(\rl_0 , A_0 )}{\partial A} = 0 .
\end{equation}

\noindent Thus, we have the equations of motion (\ref{dS/dl=0}) (at $A = A_0$) and (\ref{dS/dA=0}) and the condition of maximizing (\ref{def-l0}) (at $A = A_0$), which defines $\rl_0$.

Having written down the typical area of the spatial and diagonal triangles at which (\ref{def-l0}) has a maximum as $b^2 / 2$, we can write for the typical length scale thus determined \cite{our1}:
\begin{equation}\label{b=sqrt}                                              
b = \sqrt{ 32 G ( \eta - 5) / 3 }, \mbox{~~~ in Planck units ~} b = \sqrt{ 4 ( \eta - 5) / (3 \pi ) } .
\end{equation}

\noindent $\eta$ parameterizes the quantum extension of the theory and appears in the exponent of the volume $V_{\sigma^4}$ of the 4-simplex $\sigma^4$ in the functional measure as $V_{\sigma^4}^\eta$.

Then the task is to solve the Regge skeleton equations (in the presence of the electromagnetic field) (\ref{dS/dl=0}) and the discrete electrodynamic ones (\ref{dS/dA=0}).

As for the Regge equations, we have considered \cite{our2} a simplicial complex, in which the vertices are assigned some coordinates. This uniquely defines some piecewise affine metric. We can define the discrete analogue of the unique torsion-free metric-compatible affine connection or discrete Christoffel symbols, find the defect angles and Regge action in terms of these symbols and expand over metric variations from 4-simplex to 4-simplex. On the simplest periodic simplicial complex (with the cubic cell divided by diagonals into $4! = 24$ 4-simplices) \cite{RocWil}, in the main order of this expansion, we have the Hilbert-Einstein action in a finite-difference form,
\begin{eqnarray}\label{DM+MM}                                               
S_{\rm g} ( \rl )  = \frac{1}{16 \pi G}\sum_{\rm 4-cubes} {\cal K}^{\lambda \mu}_{~~~ \lambda \mu} \sqrt{g} , ~~~ {\cal K}^\lambda_{~ \, \mu \nu \rho} \! = \! \Delta_\nu M^\lambda_{\rho \mu} \! - \! \Delta_\rho M^\lambda_{\nu \mu} \! + \! M^\lambda_{\nu \sigma} M^\sigma_{\rho \mu} \! - \! M^\lambda_{\rho \sigma} M^\sigma_{\nu \mu} , \nonumber \\ \hspace{-10mm}
M^\lambda_{\mu \nu} = \frac{1}{2} g^{\lambda \rho} (\Delta_\nu g_{\mu \rho} + \Delta_\mu g_{\rho \nu} - \Delta_\rho g_{\mu \nu}), ~~~ \Delta_\lambda = 1 - \overline{T}_\lambda . \hspace{10mm}
\end{eqnarray}

\noindent Here $T_\lambda f(\dots , x^\lambda , \dots ) = f(\dots , x^\lambda + 1 , \dots )$ (shift operator), $f$ is a function.

\section{Electromagnetic field}\label{em-field}

As for the electromagnetic field, its simplicial form was proposed in the papers [\citen{Sor,Wein}]. In Weingarten's setting \cite{Wein}, the electromagnetic field is a 2-chain or a combination of 2-simplices (triangles). Indeed, we can decompose the electromagnetic field tensor over a set of the triangle bivectors,
\begin{equation}                                                            
F_{\lambda \mu} = \sum_{\sigma^2} c_{\sigma^2} [l_{\sigma^1_1} , l_{\sigma^1_2}]_{\lambda \mu} , ~~~ [l_{\sigma^1_1} , l_{\sigma^1_2}]_{\lambda \mu} \equiv \frac{1}{2} (l_{\sigma^1_1 \lambda } l_{\sigma^1_2 \mu } - l_{\sigma^1_1 \mu } l_{\sigma^1_2 \lambda })
\end{equation}

\noindent (symbolically), and the coefficients $c_{\sigma^2}$ can serve as variables. In Sorkin's setting \cite{Sor}, the variables are the 2-forms on the 2-simplices or electromagnetic field fluxes through the triangles or "elementary" fluxes,
\begin{equation}                                                           
\Phi_{\sigma^2} \equiv F ( [ \sigma^1_1 \sigma^1_2 ] ) = \frac{1}{2} F_{\lambda \mu} [l_{\sigma^1_1} , l_{\sigma^1_2}]^{\lambda \mu} = \frac{1}{2} \sum_{\sigma^{2 \prime}} c_{\sigma^{2 \prime}} [l_{\sigma^{1 \prime}_1} , l_{\sigma^{1 \prime}_2}]_{\lambda \mu} [l_{\sigma^1_1} , l_{\sigma^1_2}]^{\lambda \mu}
\end{equation}

\noindent - 2-form $F$ on the 2-simplex $\sigma^2 \equiv [ \sigma^1_1 \sigma^1_2 ] $ spanned by the 1-simplices (edges) $\sigma^1_1$, $\sigma^1_2$. The sets $c_{\sigma^2}$ and $\Phi_{\sigma^2}$ are in a sense dual and related by the matrix of products of bivectors of triangles.

The first pair of the simplicial Maxwell equations says that there exists a 1-form $A$ on the 1-simplices such that $F = \d A$ or the flux of $F$ through a triangle is the line integral around its boundary or a sum of the integrals along its edges oriented in the direction of passing along its boundary,
\begin{equation}\label{F=dA}                                               
F(\sigma^0_1 \sigma^0_2 \sigma^0_3 ) = A (\sigma^0_1 \sigma^0_2 ) + A (\sigma^0_2 \sigma^0_3 ) + A (\sigma^0_3 \sigma^0_1 ),
\end{equation}

\noindent using Sorkin's interpretation in terms of forms. Here $\sigma^0_1 \sigma^0_2 \sigma^0_3$ is the triangle with the vertices $\sigma^0_1$, $\sigma^0_2$, $\sigma^0_3$ and the edges $\sigma^0_1 \sigma^0_2$, $\sigma^0_2 \sigma^0_3$, $\sigma^0_3 \sigma^0_1$.

The number of these new independent variables $A (\sigma^0_1 \sigma^0_2 )$ is the number of the edges or, in our simplest periodic simplicial complex, 15 per vertex. It is still large compared to the required number of the degrees of freedom 4 per vertex for the vector potential. To concentrate on the latter degrees of freedom, we define the integral of the vector potential along an edge approximately through the half-sum of its values at the ends of the edge,
\begin{equation}\label{A=(A+A)/2}                                          
A (\sigma^0_1 \sigma^0_2 ) = \frac{1}{2} [ A_\lambda (\sigma^0_1 ) + A_\lambda (\sigma^0_2 ) ] l^\lambda_{\sigma^0_1 \sigma^0_2 }.
\end{equation}

\noindent This expression

1) is exact if a piecewise linear (inside the 4-simplices) ansatz for the vector potential field is considered;

2) generally holds in the leading order over variations of variables from vertex to vertex;

3) can be thought of as precise as a parameterization (that reduces a redundant set of abstract variables to a smaller set of other abstract variables).

4) Finally, this fits our approach, which deals in zero order with variables at the vertices.

Note that the variable $A_\lambda (\sigma^0 )$ is not a 1-form.

We can find $F_{\lambda \mu} ( \sigma^4 )$ inside the 4-simplices $\sigma^4$ of our periodic complex in terms of $A_\lambda (\sigma^0 )$. For this complex, a special notation is convenient. The vertices of each of its 4-simplices in a given cell are a certain $O$ and those obtained by subsequent applying the translation operator $T_\lambda$ along the different coordinates $x^\lambda$:
\begin{equation}                                                           
O , ~~ T_\lambda O , ~~ T_{\lambda \mu } O , ~~ T_{\lambda \mu \nu } O , ~~ T_{\lambda \mu \nu \rho } O .
\end{equation}

\noindent ($T_{\lambda \dots \mu }$ is the product $T_\lambda \dots T_\mu$.) This 4-simplex will be denoted as $[\lambda \mu \nu \rho]$. The $n$-simplices at $n < 4$ are denoted analogously; what is new is that there may be intermediate shifts along two or more coordinates simultaneously, for example, the 3-simplex obtained by shifts of $O$ by $T_\lambda$, $T_{\mu \nu }$, $T_\rho$ is denoted by $[\lambda ( \mu \nu ) \rho]$. The 4-simplex $[\lambda \mu \nu \rho]$ contains 10 2-simplices,
\begin{eqnarray}                                                           
& & [\lambda \mu ] , ~ [\lambda ( \mu \nu )] , ~ [ ( \lambda \mu ) \nu ] , ~
[\lambda ( \mu \nu \rho )] , ~ [(\lambda \mu ) ( \nu \rho )] , ~ [(\lambda \mu \nu ) \rho] , \nonumber \\ & & T_\lambda [ \mu \nu ] , ~ T_\lambda [ \mu ( \nu \rho )] , ~ T_\lambda [( \mu \nu ) \rho] , ~ T_{\lambda \mu } [ \nu \rho] .
\end{eqnarray}

\noindent On one hand, the flux of $F$ through any of these triangles $[\sigma^1_1 \sigma^1_2 ] = [(\lambda_1 \dots \lambda_k ) (\mu_1 \dots \mu_n ) ]$ is
\begin{equation}\label{Fll}                                                
F([\sigma^1_1 \sigma^1_2 ]) = \frac{1}{2} F_{\lambda \mu} l^\lambda_{\sigma^1_1} l^\mu_{\sigma^1_2} = \frac{1}{2} \sum_{\lambda = \lambda_1 , \dots , \lambda_k } \sum_{\mu = \mu_1 , \dots , \mu_n } F_{\lambda \mu} .
\end{equation}

\noindent On the other hand, it is calculated in terms of $A_\lambda (\sigma^0 )$ from (\ref{F=dA}) and (\ref{A=(A+A)/2}) giving
\begin{equation}\label{Fll=Al}                                             
2 F([\sigma^1_1 \sigma^1_2 ]) = T_{\sigma^1_1 } \Delta_{\sigma^1_1 } \sum_{\lambda = \lambda_1 , \dots , \lambda_k } A_\lambda - T_{\sigma^1_1 } T_{\sigma^1_2 } \Delta_{\sigma^1_2 } \sum_{\mu = \mu_1 , \dots , \mu_n } A_\mu .
\end{equation}

\noindent Here $\Delta_{\sigma^1 } \equiv 1 - \overline{T}_{\sigma^1 }$ (a finite difference). Comparing (\ref{Fll}) and (\ref{Fll=Al}) gives $F_{\lambda \mu}$ in the 4-simplex $[\nu \dots \rho \lambda \dots \mu \dots ]$,
\begin{equation}                                                           
F_{\lambda \mu} = T_{\nu \dots \rho } T_\lambda (\Delta_\lambda A_\mu - T_\mu \Delta_\mu A_\lambda )
\end{equation}

\noindent (the set $\{ \nu \dots \rho \}$ may be empty). We see that in the leading order over field variations from point to point, $F_{\lambda \mu}$ becomes a finite-difference form of the standard expression for the electromagnetic tensor, the same for each of the $4!=24$ 4-simplices of the 4-cubic cell. Correspondingly, in the main order over field variations, we have a finite-difference form of the standard electromagnetic action,
\begin{equation}                                                           
S_{\rm em} = - \frac{1}{16 \pi } \sum_{\rm 4-cubes} F^{\lambda \mu} F_{\lambda \mu} \sqrt{g} , ~~~ F_{\lambda \mu} = \Delta_\lambda A_\mu - \Delta_\mu A_\lambda + O([\Delta ]^2).
\end{equation}

\section{The metric ansatz}\label{ansatz}

As mentioned in Section (\ref{method}), to calculate the measure $F ( \rl )$ in the functional integral (\ref{int-exp(iS)-dmu-DOmega}), it is appropriate to use the expansion over the discrete lapse-shift functions (in which the zero order is taken here). In order not to worsen the convergence of the expansion, it is necessary to ensure that these functions are bounded. The simplest option is to work in a frame of reference close to synchronous, so that we substitute in the finite-difference form (\ref{DM+MM}) an interpolating metric with $(N, \bN ) = (1, {\bf 0})$.

In the case of the Schwarzschild black hole, a synchronous frame of reference can be realized by binding the coordinates to the set of radially freely falling particles that rest at infinity (the Lemaitre metric). In the case of the Reissner–Nordström solution, the particles at rest at infinity, that is, having the total energy $E = 1$, cannot reach the center, and the direct analogue of the Lemaitre coordinate system does not cover the neighborhood of the center. The coordinates should be tied to particles with an arbitrarily large energy. The idea is to use the energy $E$ itself as a coordinate. To this end, it is appropriate to use a general approach based on solving the Hamilton-Jacobi equation, as described, for example, in [\citen{Landau}]. For the spherically-symmetrical case of the continuum Reissner–Nordström solution,
\begin{equation}\label{RNmetric}                                           
\d s^2 = - h \d t^2 + h^{-1} \d r^2 + r^2 \d \Omega^2 , ~~~ h = 1 - \frac{r_g}{r } + \frac{q^2 }{r^2 } ,
\end{equation}

\noindent it is sufficient to consider the truncated $r, t$-part of the metric and the corresponding two-dimensional Hamilton-Jacobi equation $( \nabla_\lambda \tau_{\rm s} ) ( \nabla^\lambda \tau_{\rm s} ) = 1$,
\begin{equation}                                                           
- h^{-1} \left ( \frac{\partial \tau_{\rm s}}{\partial t} \right )^2 + h \left ( \frac{\partial \tau_{\rm s}}{\partial r} \right )^2 + 1 = 0 .
\end{equation}

\noindent It is separable, and the completely separated solution has the form
\begin{equation}\label{tau=Et+R}                                           
\tau_{\rm s} = - E t + R ( E, r ) ,
\end{equation}

\noindent so
\begin{equation}                                                           
- h^{-1} E^2 + h ( R^\prime_r )^2 + 1 = 0, R = \int^r \frac{\sqrt{ E^2 - h }}{h } \d r .
\end{equation}

\noindent Here $E$ is a constant, and it defines the reachable values of $r$ via $E^2 \geq h$. One else constant is additive to $\tau_{\rm s}$ and can be included into the definition of $\tau_{\rm s}$. The equation of motion takes the form $\tau^\prime_{{\rm s} \, E} = 0$ or
\begin{equation}\label{t+dR/dE=0}                                          
- t + R^\prime_E = 0 .
\end{equation}

\noindent Any choice of $E$ defines a certain trajectory in the $r, t$-plane, and $E$ can be chosen as a new spacelike coordinate, while $\tau_{\rm s}$ can be chosen as a new time. Eqs. (\ref{tau=Et+R}) and (\ref{t+dR/dE=0}) relate $(\tau_{\rm s} , E)$ and $(t, r)$ and allow to express $\d t, \d r$ in terms of $\d \tau_{\rm s} , \d E$ and find
\begin{equation}                                                           
\d s^2 = - \d \tau_{\rm s}^2 + (E^2 - h) \left ( R_{E^2}^{\prime \prime} \right )^2 \d E^2 + r^2 (\tau_{\rm s} , E) \d \Omega^2
\end{equation}

\noindent ($E^2 \geq h$ in the domain of definition of the new coordinates).

A triangulation can be done taking into account the coordinate lines of the considered coordinates. In particular, we can consider a location of the vertices along the coordinate $E$ in the leaves $\tau_{\rm s} = const$ with the distance between them $\Delta s = b$, which determines $\Delta E \approx b (E^2 - h)^{- 1 / 2} | R_{E^2}^{\prime \prime} |^{- 1}$. In reality, we should consider a simplicial metric close at large distances in some sense to the Reissner–Nordström one, here in a synchronous frame of reference. We use the simplest periodic simplicial complex, and the Regge action in the leading order over metric variations is a finite-difference form of the Hilbert-Einstein action (\ref{DM+MM}), where the interpolating smooth metric $g_{\lambda \mu}$ is close in some sense to the Reissner–Nordström one (in a synchronous frame). In principle, non-leading orders over metric variations could be taken into account in (\ref{DM+MM}), although this would greatly complicate the problem, in particular, the metric would have to be characterized by simplicial components additional to $g_{\lambda \mu}$ at each vertex.

But we restrict ourselves to the leading order over metric variations, and in this order the finite differences obey the same rules as the ordinary derivatives. In particular, in a finite-difference expression, we can go from a metric close to the Reissner–Nordström metric in a synchronous frame of reference to a metric close to the Reissner–Nordström metric in some other frame of reference (a kind of diffeomorphism invariance on the discrete level). Therefore, in this order, the details of the intermediate synchronous frame do not matter, only its existence is important. As a new metric, we can consider the Painlev\'{e}-Gullstrand type metric \cite{Painleve,Gullstrand},
\begin{equation}\label{PGmetric}                                           
\d s^2 = - \d \tau_{\rm PG}^2 + (\d r - \sqrt{ 1 - h } \d \tau_{\rm PG} )^2 + r^2 \d \Omega^2 , ~~~ \tau_{\rm PG} = t - \int^r \frac{ \sqrt{1 - h}}{ h } \d r ,
\end{equation}

\noindent and the Kerr-Schild metric \cite{Kerr,Kerr1},
\begin{equation}\label{KSmetric}                                           
\d s^2 = - \d \tau^2 + \d r^2 +r^2 \d \Omega^2 + ( 1 - h ) (\d \tau - \d r )^2 , ~~~ \tau = t - \int^r \frac{ 1 - h }{ h } \d r .
\end{equation}

\noindent The Painlev\'{e}-Gullstrand coordinate system is undefined in the neighborhood of the center, where $h > 1$, and the Kerr-Schild coordinates are devoid of this disadvantage.

In addition, in the purely Schwarzschild case, when both coordinate systems are well-defined, the results of our paper \cite{Kha1} obtained using the Painlev\'{e}-Gullstrand coordinates are reproduced here using the Kerr-Schild coordinates. Thus, we are trying to construct a discrete Reissner–Nordström solution in an analog of the Kerr-Schild coordinates.

\section{The discrete solution}\label{solution}

We can write (in many ways) any metric $g_{\lambda \mu}$ in the form
\begin{equation}                                                           
g_{\lambda \mu} = g_{(0) \lambda \mu} + l_\lambda l_\mu , ~~~ l^\lambda l_\lambda = 0 , ~~~ l^\lambda = g_{(0) }^{ \lambda \mu} l_\mu .
\end{equation}

\noindent (If $g_{(0) \lambda \mu}$ is a flat metric, then we can speak of the algebraically special metric $g_{\lambda \mu}$.) Then the Riemann tensor $R^\lambda_{\mu \nu \rho}$ for $g_{\lambda \mu}$ is related to the tensor $R^\lambda_{( 0 ) \mu \nu \rho}$ for $g_{(0) \lambda \mu}$ as
\begin{equation}                                                           
R_{ \mu \nu \rho}^\lambda = R_{(0) \mu \nu \rho}^\lambda + \nabla_\nu C^\lambda_{\mu \rho } - \nabla_\rho C^\lambda_{\mu \nu } + C^\lambda_{\sigma \nu} C^\sigma_{\mu \rho} - C^\lambda_{\sigma \rho} C^\sigma_{\mu \nu} ,
\end{equation}

\noindent where $\nabla_\lambda$ means the covariant derivative with respect to $g_{(0) \lambda \mu}$ and
\begin{equation}                                                           
C^\lambda_{\mu \nu } = \frac{1}{2} \left [ \nabla_\mu (l_\nu l^\lambda ) + \nabla_\nu (l_\mu l^\lambda ) - \nabla^\lambda ( l_\mu l_\nu ) + l^\lambda l^\sigma \nabla_\sigma (l_\mu l_\nu ) \right ] ~~~ ( C^\lambda_{\mu \lambda } = 0 ) .
\end{equation}

\noindent The Ricci tensor is
\begin{equation}                                                           
R_{\lambda \mu } = R_{ (0) \lambda \mu } + \nabla_\nu C^\nu_{\lambda \mu } - C^\nu_{\rho \mu} C^\rho_{\lambda \nu} .
\end{equation}

\noindent For $g_{(0) \lambda \mu} = \eta_{\lambda \mu} \equiv \mbox{diag} (-1, 1, 1, 1)$, $\nabla_\lambda = \partial_\lambda$ and in the order $[l]^2$, this reads
\begin{eqnarray}                                                           
& & R_{0 0} = - \frac{1}{2} \bpart^2 ( l_0^2 ) , \nonumber \\
& & R_{0 k} = - \frac{1}{2} \bpart^2 ( l_0 l_k ) + \frac{1}{2} \partial_k \partial_m (l_0 l_m ) , \nonumber \\
& & R_{k l} = - \frac{1}{2} \bpart^2 ( l_k l_l ) + \frac{1}{2} \partial_k \partial_m (l_l l_m ) + \frac{1}{2} \partial_l \partial_m (l_k l_m ) .
\end{eqnarray}

\noindent We denote
\begin{equation}                                                           
l_\lambda = - l_0 n_\lambda , ~~~ n_\lambda = ( -1 , \bn ) = ( -1 , n_1 , n_2 , n_3 ) , ~~~ \bn = \bx r^{- 1} .
\end{equation}

\noindent As is known, the Reissner - Nordström electromagnetic potential can be written as $\tl_0 \tl_\lambda$, where $\tl_\lambda$ is proportional to $l_\lambda$,
\begin{equation}                                                           
A_\lambda = - \phi n_\lambda \equiv \tl_0 \tl_\lambda , ~~~ l_\lambda = \kappa ( \bx ) \tl_\lambda
\end{equation}

\noindent ($\kappa ( \bx )$ is a function), and is itself a solution of the gravity equations for $l_\lambda$ in empty spacetime, in particular \cite{Chandra},
\begin{equation}\label{ldl=l}                                              
\tl^\nu \partial _\nu \tl_\lambda = \psi (\bx ) \tl_\lambda
\end{equation}

\noindent ($\psi ( \bx )$ is a function). Raising the indices of the electromagnetic tensor turns out to be simple,
\begin{equation}                                                           
F^{\lambda \mu} = \eta^{\lambda \nu} \eta^{\mu \rho} F_{\nu \rho}, ~~~ (F_{\lambda \mu} = \partial_\lambda A_\mu  - \partial_\mu A_\lambda )
\end{equation}

\noindent (equation (\ref{ldl=l}) is used). For its components, we have
\begin{equation}                                                           
F_{0 k} = - \partial_k \phi, F_{k l} = n_k \partial_l \phi - n_l \partial_k \phi ,
\end{equation}

\noindent and for its divergence,
\begin{equation}                                                           
F^{\lambda \mu }{}_{; \lambda } = (- g)^{- 1 / 2} \frac{\partial}{\partial x^\lambda } [(- g)^{1 / 2} F^{\lambda \mu } ] = \frac{\partial}{\partial x^\lambda } F^{\lambda \mu } ~~~ ( g = -1 ) ,
\end{equation}

\noindent we obtain
\begin{eqnarray}                                                           
& & F^{\lambda 0 }{}_{; \lambda } = - \bpart^2 ( \tl_0^2 ) = - \bpart^2 \phi , \nonumber \\
& & F^{\lambda k }{}_{; \lambda } = \bpart^2 ( \tl_0 \tl_k ) - \bpart_k \bpart_m ( \tl_0 \tl_m ) = - \bpart^2 ( \phi n_k ) + \bpart_k \bpart_m ( \phi n_m ) .
\end{eqnarray}

\noindent For the electromagnetic stress-energy tensor,
\begin{equation}                                                           
4 \pi T_{\lambda \mu} = F_{\lambda \nu } F_\mu {}^\nu - \frac{1}{4} g_{\lambda \mu } F_{\nu \rho } F^{\nu \rho } ,
\end{equation}

\noindent we have
\begin{eqnarray}                                                           
& & 4 \pi T_{0 0} = ( \partial_k \phi )^2 - \frac{1}{2} (n_k \partial_k \phi )^2 , \nonumber \\
& & 4 \pi T_{0 k} = n_k ( \partial_m \phi )^2 - ( \partial_k \phi ) ( n_m \partial_m \phi ) , \nonumber \\
& & 4 \pi T_{k l} = n_k n_l ( \partial_m \phi )^2 - ( n_m \partial_m \phi ) ( n_k \partial_l \phi + n_l \partial_k \phi ) + \frac{1}{2} ( n_m \partial_m \phi )^2 \delta_{k l} .
\end{eqnarray}

If $\phi$ is a pure function of $r$, then we have
\begin{equation}                                                           
F^{\lambda k }{}_{; \lambda } = 0 , ~~~ R_{0 k} = 0 , ~~~ T_{0 k} = 0
\end{equation}

\noindent identically (that is, spatial components of the Maxwell equations and space-time components of the Einstein equations are satisfied). The spatial-spatial components of the Einstein equations $R_{k l} = 4 \pi T_{k l}$ are satisfied identically, if the solution of the 00-component $R_{0 0} = 4 \pi T_{0 0}$ of the Einstein equations and 0-component $F^{\lambda 0 }{}_{; \lambda } = 0$ of the Maxwell equations is substituted. This pair of equations takes the form
\begin{equation}\label{cont}                                               
\left. \begin{array}{rcl} \bpart^2 (l_0^2 ) & = & - 4 G \left [ ( \partial_k \phi )^2 - \frac{1}{2} ( n_k \partial_k \phi )^2 \right ] \\ \bpart^2 \phi & = & 0 \end{array} \right \} \mbox{ at } \bx \neq {\bf 0} .
\end{equation}

In finite differences, we have
\begin{eqnarray}\label{discr-00}                                        
& & \sum^3_{j = 1} \bDelta_j \Delta_j ( l_0^2 ) = 4 G \left [ \sum^3_{j = 1} \left ( \Delta_j \phi \right )^2 - \frac{1}{2} \left ( \sum^3_{j = 1} n_j \Delta_j \phi \right )^2 \right ] \mbox{ for } \bx \neq 0 , \\ & & \label{discr-0} \sum^3_{j = 1} \bDelta_j \Delta_j \phi = 0 \mbox{ for } \bx \neq {\bf 0} .
\end{eqnarray}

In the continuum (\ref{cont}), we had
\begin{equation}                                                           
\phi = \frac{Q}{r} , ~~~ ( \partial_k \phi )^2 - \frac{1}{2} ( n_k \partial_k \phi )^2 = \frac{Q^2 }{ 2 r^4 } , ~~~ l_0^2 = \frac{ r_g }{ r } - \frac{ q^2 }{ r^2 } , ~~~ q^2 \equiv G Q^2 .
\end{equation}

In the discrete case, it is convenient to write $l_0^2$ as the sum of the general solution $( l_0^2 )_g$ of the homogeneous equation and the particular solution $( l_0^2 )_{em}$ of the inhomogeneous one simultaneously continuing the pair of equations (\ref{discr-00}, \ref{discr-0}) to the vertex $\bx = {\bf 0}$ and parameterizing this continuation by two constants $C_g = 4 \pi b^2 r_g$ and $C_{em} = 4 \pi b^2 Q$, which we include in $( l_0^2 )_g$ and $\phi$, respectively.
\begin{eqnarray}\label{discr-l0em}                                      
& & \sum^3_{j = 1} \bDelta_j \Delta_j ( l_0^2 )_{em} = 4 G \left [ \sum^3_{j = 1} \left ( \Delta_j \phi \right )^2 - \frac{1}{2} \left ( \sum^3_{j = 1} n_j \Delta_j \phi \right )^2 \right ] , \\ & & \label{discr-l0g} \sum^3_{j = 1} \bDelta_j \Delta_j ( l_0^2 )_g = \left \{ \begin{array}{rl} 0 & \mbox{at } \bx \neq 0  \\ C_g & \mbox{at } \bx = 0 , \end{array} \right. \\ & & \label{discr-phi} \sum^3_{j = 1} \bDelta_j \Delta_j \phi = \left \{ \begin{array}{rl} 0 & \mbox{at } \bx \neq 0  \\ C_{em} & \mbox{at } \bx = 0 . \end{array} \right.
\end{eqnarray}

\noindent Here $( l_0^2 )_g \equiv l_{0 g}^2$ is a version of $l_0^2$ itself (for $q = 0$), and $( l_0^2 )_{em}$ is just an additional term in $l_0^2$.

As for $n_k = x_k r^{ - 1 }$, this is an "external" explicit coordinate function that changes abruptly in the vicinity of $\bx = {\bf 0}$, and it seems appropriate to express it in terms of the already available field functions for consistency when passing to the discrete case. To this end, note that $n_k$ in the continuum arises when any pure function of $r$, $\chi (r )$, is differentiated by $\partial_k$, and then $f n_k = \partial_k \chi$, where $f^2 = ( \partial_k \chi )^2$. Also note that $n_k$ appears as a factor of $f = \tl_0$ or $l_{0 g}$, which differs from $\tl_0$ by a constant factor. Hence, a natural discrete definition of $n_k$ could be through the solution of a Hamilton-Jacobi-type equation,
\begin{equation}\label{discr-chi-n}                                        
b^{ - 2 } \sum^3_{j = 1} \left ( \Delta_j \chi \right )^2 = l_{0 g}^2 , ~~~ n_j = b^{ - 1 } l_{0 g}^{-1 } \Delta_j \chi .
\end{equation}

\noindent We consider this $\chi$ in our previous paper \cite{Kha1}, and the inverse of the Laplacian giving the present $l_{0 g}^2$ and $\phi$ was used there,
\begin{equation}\label{discr-l0g-phi}                                      
\frac{l_{0 g}^2}{r_g} = \frac{\phi}{Q} = \int^{ \pi / b }_{\! \! - \pi / b } \int^{ \pi / b }_{\! \! - \pi / b } \int^{ \pi / b }_{\! \! - \pi / b } \frac{\d^3 \bp}{(2 \pi )^3} \frac{ \pi b^2 \exp (i \bp \, \bx )}{\sum_k \sin^2 (p_k b / 2 )} .
\end{equation}

\noindent In particular, we take the continuum value $\chi ( \bx ) = 2 \sqrt{ r_g r }$ at $r \geq b$, and we find $\chi ( {\bf 0 } )$ from (\ref{discr-chi-n}) at $\bx = {\bf 0}$,
\begin{equation}                                                           
3 b^{-2} [ \chi ({\bf 0}) - \chi ( -b, 0, 0) ]^2 = l_{0 g}^2 ({\bf 0}) = 1.05 \pi r_g b^{-1} , ~~~ \chi ( {\bf 0} ) = 0.95 \sqrt{r_g b }
\end{equation}

\noindent ($l_{0 g}^2 ({\bf 0}) r_g^{-1}$ is the value of the table integral (\ref{discr-l0g-phi}) at $\bx = {\bf 0}$). With this $\chi ({\bf 0})$,
\begin{eqnarray}\label{1.39/b}                                             
l_{0 g}^2 (b, 0, 0) & = & b^{-2} [\chi (b, 0, 0) - \chi ({\bf 0})]^2 + b^{-2} [\chi (b, 0, 0) - \chi (b, -b, 0)]^2 \nonumber \\ & & + b^{-2} [\chi (b, 0, 0) - \chi (b, 0, -b)]^2 = 1.39 r_g b^{- 1} .
\end{eqnarray}

\noindent At the same time, the exact (\ref{discr-l0g-phi}) at $\bx = (\pm b , 0, 0), (0, \pm b, 0), (0, 0, \pm b)$ is $1.19 b^{-1}$. Found (\ref{1.39/b}) through $\chi$ is 20\% more than this exact $1.19 r_g b^{-1}$ (and the continuum value $r_g b^{-1}$, on the contrary, is 20\% less). This way of estimating through $\chi$ leads to equal values $n_k$ at $\bx = {\bf 0}$,
\begin{equation}                                                           
n_k = \frac{1}{b l_{0 g}} \Delta_k \chi ( 0 ) = - \frac{1}{ \sqrt{3 }} , ~~~ k = 1, 2, 3 .
\end{equation}

Then we can estimate the RHS of (\ref{discr-l0em}) at $\bx = {\bf 0}$,
\begin{equation}                                                           
\sum^3_{j = 1} \left ( \Delta_j \phi ({\bf 0}) \right )^2 - \frac{1}{2} \left ( \sum^3_{j = 1} n_j \Delta_j \phi ({\bf 0}) \right )^2 = \frac{2}{3} \frac{ \pi^2 Q^2 }{ b^2 } .
\end{equation}

\noindent (the estimate of $\Delta_j \phi ({\bf 0})$ from the integral (\ref{discr-phi}) can be done exactly as $2 \pi Q (3 b)^{-1}$).
Again, adopting the continuum value for $( l_0^2 )_{em}$ at $r \geq b$,
\begin{equation}                                                           
( l_0^2 )_{em} ( \bx ) = - \frac{ q^2 }{ r^2 } , ~~~ r \geq b ,
\end{equation}

\noindent and taking $( l_0^2 )_{em} ( {\bf 0} )$ as an unknown value, we can write (\ref{discr-l0em}) at $\bx = {\bf 0}$ and find this unknown, resulting in
\begin{equation}\label{l02(0)}                                             
l_0^2 ( {\bf 0} ) = 1.05 \pi \frac{ r_g }{ b } + ( l_0^2 )_{em} ( {\bf 0} ) = 1.05 \pi \frac{ r_g }{ b } + \frac{ 4 \pi^2 - 9 }{ 9 } \frac{ q^2 }{ b^2 } .
\end{equation}

\noindent An interesting feature of this result is the positivity of $( l_0^2 )_{em} ( \bx )$ at $\bx = {\bf 0}$, despite its growing negativity when approaching the center up to the vertices closest to the center. The origin of this result can be seen from the considered equation (\ref{discr-l0em}). On the LHS of the equation at the center, we have $6 [ (l_0^2 )_{em} ( {\bf 0} ) - (l_0^2 )_{em} (b, 0, 0) ]$ (taking into account the symmetry between the vertices $(\pm b , 0, 0), (0, \pm b, 0), (0, 0, \pm b)$), and this is positive due to the positiveness of the stress-energy component on the RHS. Thus, $(l_0^2 )_{em} ( {\bf 0} )$ is shifted positively compared to $(l_0^2 )_{em} (b, 0, 0)$, and this shift can be large enough if the stress-energy component is numerically large, as we just have obtained.

\section{Riemann tensor}\label{riemann}

To evaluate the curvature, we have in the order $[l]^2$ for the Riemann tensor,
\begin{equation}                                                           
R_{\lambda \mu \nu \rho} = \frac{1}{2} \left [ \partial_\mu \partial_\nu \left ( l_\lambda l_\rho \right ) + \partial_\lambda \partial_\rho \left ( l_\mu l_\nu \right ) - \partial_\mu \partial_\rho \left ( l_\lambda l_\nu \right ) - \partial_\lambda \partial_\nu \left ( l_\mu l_\rho \right ) \right ]
\end{equation}

\noindent (in the continuum) or, separating the temporal and spatial indices,
\begin{eqnarray}\label{Rieman}                                             
& & R_{0 k 0 l} = - \frac{1}{2} \partial_k \partial_l ( l_0^2 ) , \nonumber \\ & & R_{0 k l m} = \frac{1}{2} \partial_k [ \partial_l (l_0 l_m) - \partial_m (l_0 l_l)] , \nonumber \\ & & R_{k l m n} = \frac{1}{2} [ \partial_l \partial_m ( l_k l_n ) + \partial_k \partial_n ( l_l l_m ) - \partial_l \partial_n ( l_k l_m ) - \partial_k \partial_m ( l_l l_n )] .
\end{eqnarray}

\noindent Here
\begin{equation}                                                           
l_k l_l = l_0^2 n_k n_l = z ( \partial_k \chi ) (\partial_l \chi ) , ~~~ z = \frac{ l_0^2 }{ l_{0 g}^2 } = 1 + \frac{ (l_0^2 )_{em} }{ l_{0 g}^2 } .
\end{equation}

\noindent We denote $z_k \equiv \partial_k z$, $z_{k l} \equiv \partial_k \partial_l z$, $\chi_k \equiv \partial_k \chi$, $\chi_{k l} \equiv \partial_k \partial_l \chi$ and find
\begin{eqnarray}\label{Rklmn}                                              
& & R_{k l m n} = z (\chi_{k m} \chi_{l n} - \chi_{k n} \chi_{l m}) + \frac{1}{2} [z_l (\chi_{k m} \chi_n - \chi_{k n} \chi_m) + z_m (\chi_{n l} \chi_k - \chi_{n k} \chi_l) \nonumber \\ & & + z_k (\chi_{l n} \chi_m - \chi_{l m} \chi_n) + z_n (\chi_{m k} \chi_l - \chi_{m l} \chi_k)] + \frac{1}{2} [ z_{l m} \chi_k \chi_n + z_{k n} \chi_l \chi_m \nonumber \\ & & - z_{l n} \chi_k \chi_m - z_{k m} \chi_l \chi_n ] .
\end{eqnarray}

\noindent In the discrete setting, we substitute $\partial_k$ by $b^{-1} \Delta_k$. Such a substitution is not unique, but this non-uniqueness is at the level of higher orders in $\Delta$; their inclusion makes sense if we also take them into account when passing from the original Regge action to the finite-difference action (\ref{DM+MM}). Here we should provide a consistence with the considered field equations, in which we use the simplest Hermitian form of the difference Laplacian (\ref{discr-00},\ref{discr-0}). The field equations are combinations of the components of the Ricci tensor and, therefore, of the Riemann tensor. To get such a Laplacian, the second derivative in $R_{0k0l}$ (\ref{Rieman}) should be analogously substituted by such an operator, $\partial_l \partial_k \to - b^{-2} (\bDelta_l \Delta_k + \Delta_l \bDelta_k ) / 2$. We also take the following metric functions,
\begin{eqnarray}                                                           
& & \chi ( {\bf 0} ) = 0.95 \sqrt{r_g b} , ~~~ \chi ( \bx ) = 2 \sqrt{r_g r} , ~ r \geq b ; ~~~ l_{0 g}^2 ( {\bf 0} ) = 1.05 \pi \frac{r_g }{ b } , \nonumber \\ & & l_{0 g}^2 (\pm b , 0, 0) = l_{0 g}^2 (0, \pm b, 0) = l_{0 g}^2 (0, 0, \pm b) = 1.19 \frac{r_g }{ b } ( = l_{0 g}^2 ( {\bf 0} ) - \frac{2 \pi r_g}{ 3 b } ) , \nonumber \\ & & l_{0 g}^2 ( \bx ) = \frac{r_g }{ r } , ~ r > b ; ~~~ ( l_0^2 )_{em} ( {\bf 0} ) = \frac{ 4 \pi^2 - 9 }{ 9 } \frac{ q^2 }{ b^2 } , ~~~ ( l_0^2 )_{em} ( \bx ) = - \frac{ q^2 }{ r^2 } , ~ r \geq b \hspace{5mm}
\end{eqnarray}

\noindent ($\chi $ and $l_{0 g}^2$, there $\vf^2$, were considered in our previous paper \cite{Kha1}). At the center, due to the symmetry between the three spatial directions, it is sufficient to know a few typical finite differences to be substituted into the curvature (\ref{Rklmn}),
\begin{eqnarray}                                                           
& & \chi_1 ( = b^{-1} \Delta_1 \chi ({\bf 0}) ) = - 1.05 \sqrt{\frac{r_g }{b }} , ~~~ \chi_{1 1} = - 0.22 \sqrt{\frac{r_g }{ b^3 }} , ~~~ \chi_{1 2} = - 0.67 \sqrt{\frac{r_g }{ b^3 }} , \nonumber \\ & & z = 1 + 1.03 \frac{ q^2 }{ r_g b }, ~~~ z_1 = 1.87 \frac{ q^2 }{ r_g b^2 } , ~~~ z_{1 1} = 2.21 \frac{ q^2 }{ r_g b^3 } , ~~~ z_{1 2} = 2.00 \frac{ q^2 }{ r_g b^3 } .
\end{eqnarray}

\noindent This gives the typical curvature components,
\begin{eqnarray}                                                           
R_{1212} & = & z (\chi_{11}^2 - \chi_{12}^2) + 2 z_1 \chi_1 (\chi_{11} - \chi_{12}) + (z_{12} - z_{11}) \chi_1^2 \nonumber \\ & = & - 0.40 \frac{r_g }{ b^3 } - 2.41 \frac{ q^2 }{ b^4 } , \nonumber \\ R_{1213} & = & z \chi_{12} (\chi_{11} - \chi_{12}) + z_1 \chi_1 (\chi_{11} - \chi_{12}) + \frac{1}{2} ( z_{12} - z_{11} ) \chi_1^2 \nonumber \\ & = & - 0.30 \frac{r_g }{ b^3 } - 1.31 \frac{ q^2 }{ b^4 } .
\end{eqnarray}

\noindent $R_{0 k l m}$ is zero in the considered order. For the rest of the typical space-time components, we have
\begin{eqnarray}                                                           
& & R_{0101} = \frac{1}{2 b^2 }\bDelta_1 \Delta_1 ( l_0^2 ) = \frac{ 2 \pi }{ 3 } \frac{r_g }{ b^3 } + \frac{ 4 \pi^2 }{ 9 } \frac{ q^2 }{ b^4 } = 2.09 \frac{r_g }{ b^3 } + 4.39 \frac{ q^2 }{ b^4 } , \nonumber \\ & & R_{0102} = \frac{1}{4 b^2 } ( \bDelta_1 \Delta_2 + \bDelta_2 \Delta_1 ) ( l_0^2 ) = 0.80 \frac{r_g }{ b^3 } + \frac{ 8 \pi^2 + 9 }{ 36 } \frac{ q^2 }{ b^4 } = 0.80 \frac{r_g }{ b^3 } + 2.44 \frac{ q^2 }{ b^4 } . \hspace{5mm}
\end{eqnarray}

\noindent So the Kretschmann scalar at the center can be estimated,
\begin{eqnarray}\label{RRdiscr}                                            
R_{\lambda \mu \nu \rho} R^{\lambda \mu \nu \rho} ({\bf 0}) & = & 12 (R_{1212} )^2 + 24 (R_{1213} )^2 + 12 (R_{0101} )^2 + 24 (R_{0102} )^2 \nonumber \\ & = & 71.9 \frac{r_g^2}{b^6} + 355.9 \frac{ r_g q^2 }{ b^7 } + 485.0 \frac{ q^4 }{ b^8 } .
\end{eqnarray}

The continuum value for the Kretschmann scalar \cite{ConnHenry}
\begin{equation}                                                           
R_{\lambda \mu \nu \rho} R^{\lambda \mu \nu \rho} = 12 \frac{r_g^2}{r^6} - 48 \frac{ r_g q^2 }{ r^7 } + 42 \frac{ q^4 }{ r^8 }
\end{equation}

\noindent is, of course, undefined at $r = 0$; being cut off at some small $r_{\rm c}$, it has the opposite sign of the mixed term $r_g q^2$ to that which we have obtained for such a term at the center in the discrete setting. This is due to the above considered sharp change in the sign of the electromagnetic term in the metric upon going to the center from the nearest vertices. Also note that the discrete Kretschmann scalar at the center (\ref{RRdiscr}) in terms of $r_{\rm c} = 0.742 b$ takes the form
\begin{equation}                                                           
R_{\lambda \mu \nu \rho} R^{\lambda \mu \nu \rho} ({\bf 0})  = 12.0 \frac{r_g^2}{r_{\rm c}^6} + 44.1 \frac{ r_g q^2 }{ r_{\rm c}^7 } + 44.6 \frac{ q^4 }{ r_{\rm c}^8 } .
\end{equation}

\noindent That is, the term $r_g^2$ here is such a continuum term, cut off at $r = r_{\rm c}$; in addition, the correct scales of the $q$-dependent terms, but not the sign of the term $r_g q^2$, can also be obtained by cut off at this $r$.

\section{Conclusion}

A new feature of the considered problem is necessity of incorporating electromagnetic field as a particular case of matter fields. For this, we consider the implementation of the Sorkin-Weingarten simplicial electrodynamics. In particular, in the leading order over field variations, this amounts to the finite difference Maxwell equations. Although, of course, the simplicial electrodynamics is not simply a naive finite-difference discretization of the usual continuum electrodynamics, but retains geometric features of the continuum theory.

The calculations are performed in the leading order over metric/field variations and continued for an estimate to the center. We estimate metric (and field) and curvature at the center. The result for the uncharged case $q=0$ reproduces that of our paper for the Schwarzschild case. This would be expected for a continuous theory, but is not a rule for a discrete theory, since different coordinate systems were used, roughly speaking, of the Painlevé-Gullstrand type with an unknown function $f = \sqrt{1 - h}$ (\ref{PGmetric}) in the Schwarzschild case and the Kerr-Schild type with an unknown function $1 - h = f^2$ (\ref{KSmetric}) in the Reissner–Nordström case. Differential identities valid in the continuum theory, for example, $\partial_k ( f^2 ) = 2 f \partial_k f$, cease to be identities when derivatives are replaced by finite differences. However, here, to simplify the estimate, we first performed transformations at the continuum level, and only then replaced the derivatives with finite differences. This gives correspondence with our previous paper on the Schwarzschild black hole.

An interesting feature of the discrete solution is the change in the sign of the electromagnetic term in the metric at the center compared to the sign of this term at the vertices nearest to the center. In the Reissner–Nordström metric (\ref{RNmetric}), the effective $ - g_{00} ( {\bf 0} ) = h ( {\bf 0} ) = 1 - l_0^2 ( {\bf 0} )$, where $l_0^2 ( {\bf 0} )$ (\ref{l02(0)}) contains the electromagnetic and $r_g$ terms of the same sign.

\section*{Acknowledgments}

The present work was supported by the Ministry of Education and Science of the Russian Federation.

\end{document}